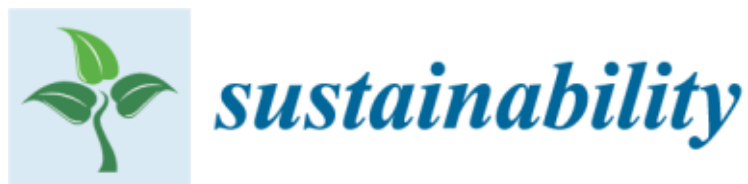

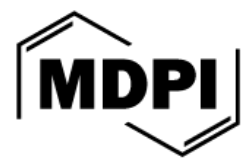

# Green Transformational Leadership, Ethical Climate, and Sustainable Nursing Practices: Evidence from the Healthcare Sector

**Thabit Atobishi [1,*] and Saeed Nosratabadi [2]**

[1] Health Information Management and Technology Department, King Faisal University, Al Ahsa 31982, Saudi Arabia; tatubaishi@kfu.edu.sa
[2] Rawls College of Business, Texas Tech University; Lubbock TX 79401, United States; snosrata@ttu.edu
* Correspondence: tatubaishi@kfu.edu.sa

**Abstract**

The healthcare sector contributes approximately 4.4% of global greenhouse gas emissions, yet research on the organizational determinants of sustainable behaviors among healthcare workers remains limited. This study examines how green transformational leadership and ethical climate influence sustainable clinical behaviors among registered nurses, with green psychological climate as a mediator and perceived organizational hypocrisy as a moderator. Data were collected from 760 nurses across 11 public and private hospitals in Jordan using a cross-sectional survey design. Structural equation modeling with bootstrapping was employed to test the hypothesized relationships. Results revealed that both green transformational leadership ($\beta = 0.215$, $p < 0.001$) and ethical climate ($\beta = 0.161$, $p < 0.001$) positively predicted sustainable clinical behaviors. Green psychological climate partially mediated both relationships. Perceived organizational hypocrisy significantly weakened the positive effects of green transformational leadership ($\beta = -0.153$, $p < 0.001$) and ethical climate ($\beta = -0.065$, $p < 0.01$) on sustainable behaviors. The model explained 35.7% of the variance in sustainable clinical behaviors. These findings highlight that fostering sustainability in healthcare requires not only supportive leadership and ethical organizational environments but also authenticity and consistency between stated values and actual practices. The study extends green transformational leadership theory to healthcare settings, integrates ethical climate research with environmental sustainability, and introduces perceived organizational hypocrisy as a critical boundary condition. Practical implications for healthcare administrators seeking to reduce their environmental footprint are discussed.

## 1. Introduction

The healthcare sector faces a profound ethical and operational paradox where the very act of preserving human health contributes significantly to environmental degradation. Recent global assessments indicate that the healthcare industry is responsible for approximately 4.4% of global net greenhouse gas emissions, a footprint comparable to the fifth-largest emitting country in the world [1]. This environmental burden arises from high energy consumption, intensive resource usage, and the generation of massive quantities of hazardous and non-hazardous waste. Consequently, there is an urgent call in the recent literature to transition from traditional care models to environmentally sustainable healthcare systems [2]. While technological and infrastructural changes are essential for addressing the largest emission sources, emerging evidence demonstrates that a substantial share of the environmental burden at the facility level is determined by the behavioral decisions of clinical staff. The World Health Organization estimates that approximately 85% of healthcare waste is general, non-hazardous material that could be recycled or safely landfilled, yet improper segregation by clinical staff frequently routes this waste to energy-intensive incineration and specialized treatment designed for the hazardous 15% [3]. Research shows that 30 to 90% of operating room waste is incorrectly classified as hazardous due to behavioral rather than material factors [4], and behavioral interventions targeting clinical staff have produced 41 to 59% reductions in regulated medical waste volumes and up to 85% reductions in associated carbon emissions without requiring infrastructure investment [4, 5]. With approximately 29.8 million nurses worldwide, comprising nearly 60% of all health professionals [6], the aggregate impact of nurses' daily clinical decisions regarding supply use, waste segregation, and resource conservation represents a systemically significant and modifiable component of healthcare's environmental footprint. Understanding the organizational factors that shape these behaviors is therefore not a peripheral concern but a question with direct implications for the scalability of healthcare decarbonization efforts.

Despite the critical role of nurses, research on the micro-foundations of sustainability within healthcare remains fragmented. In this study, sustainable clinical behaviors refer to environmentally responsible actions performed by nurses within their clinical work context, encompassing a spectrum from discretionary resource conservation (e.g., reducing energy and supply use) and proper waste management to voluntary, proactive initiatives such as advocating for green practices and encouraging colleagues to adopt sustainable routines. Current literature suggests that for nurses to engage in such behaviors, they require a supportive organizational context driven by specific leadership styles and ethical standards [7]. Notably, the existence of formal sustainability protocols in hospitals does not render leadership influence unnecessary; research consistently shows that compliance with environmental policies varies substantially, with 30 to 90% of operating room waste improperly segregated despite existing guidelines [4], and that behavioral gains from training interventions erode without sustained organizational reinforcement [8]. Green transformational leadership has been identified as a potent antecedent for bridging this gap between policy and practice. Distinct from general leadership, green transformational leaders motivate followers to achieve environmental goals by articulating a clear green vision, modeling sustainable values, and stimulating employees to think creatively about ecological problems [9]. In the high-pressure environment of a hospital, such leadership is necessary not only to enhance compliance with existing green protocols but also to motivate the discretionary and proactive environmental behaviors that formal rules alone cannot prescribe.

While positive antecedents are documented in general organizational contexts, the literature has largely overlooked the potential negative influences that may undermine sustainability efforts. As hospitals increasingly market their green credentials, there is a risk of perceived organizational hypocrisy, where employees observe a disconnect between the organization's stated values and its actual practices [10]. If nurses perceive that their leaders advocate for sustainability but fail to allocate resources or personally adhere to green protocols, cynicism may ensue. Such perceived hypocrisy can act as a critical boundary condition, potentially neutralizing the positive effects of leadership and ethical climates on sustainable behaviors [11]. This represents a significant gap in the literature, as the interplay between positive organizational antecedents and perceptions of hypocrisy has not been examined in healthcare sustainability research.

This study addresses these gaps by investigating the organizational determinants of sustainable clinical behaviors among registered nurses in Jordanian hospitals. Specifically, the study examines the influence of green transformational leadership, ethical climate, and green psychological climate on nurses' sustainable clinical behaviors, while also considering the potential undermining role of perceived organizational hypocrisy. By situating this research in Jordan, a developing nation facing acute resource scarcity, this study responds to calls for empirical evidence from non-Western healthcare contexts where the imperative for conservation is both economic and ecological [12].

This study offers three primary contributions to the literature. First, it extends the green transformational leadership literature by applying it to the healthcare sector, an underexplored context despite its significant environmental impact. Second, it integrates ethical climate research with environmental sustainability, responding to calls for examining how organizational ethics relate to pro-environmental outcomes [13]. Third, it introduces perceived organizational hypocrisy as a boundary condition in healthcare sustainability research, addressing the "dark side" of organizational perceptions that has been largely neglected in prior studies.

## 2. Literature Review and Hypotheses Development

### *2.1. Theoretical Background*

This study draws on social information processing theory and social learning theory to explain how organizational factors influence nurses' sustainable clinical behaviors. Social information processing theory posits that employees rely on cues from their immediate social environment to interpret situations and guide their attitudes and behaviors [14]. In the context of sustainability, nurses observe and interpret signals from their leaders and the broader organizational climate to determine appropriate environmental behaviors. Social learning theory further suggests that individuals learn behaviors by observing and imitating role models, particularly those in positions of authority [15]. When nurse managers demonstrate commitment to environmental sustainability, subordinates are more likely to adopt similar behaviors. Together, these theories provide a foundation for understanding how green transformational leadership, ethical climate, and green psychological climate shape nurses' engagement in sustainable clinical practices.

In this study, green transformational leadership and ethical climate are conceptualized as complementary but theoretically distinct organizational antecedents. Green transformational leadership represents a person-based influence mechanism rooted in social learning theory, whereby nurses learn sustainable behaviors by observing and imitating their immediate supervisors [15]. Ethical climate, by contrast, represents a context-based influence mechanism grounded in social information processing theory, reflecting the shared normative expectations embedded in organizational policies, procedures, and practices [14, 16]. While a green transformational leader provides a specific role model for environmental behavior, ethical climate establishes the broader moral infrastructure within which sustainability can be perceived as consistent with professional obligations. Together, they capture both the interpersonal (leader-follower) and institutional (organization-member) pathways through which organizational factors shape nurses' sustainable clinical behaviors.

Each of the multi-dimensional constructs in this study is conceptualized as a second-order factor comprising theoretically coherent sub-dimensions. Ethical climate encom-passes caring, law and code, and rules dimensions [16], which collectively represent the normative structure of the organization; social information processing theory applies uniformly across these dimensions because nurses draw behavioral cues from all three forms of ethical signaling, whether the signal concerns stakeholder welfare, professional standards, or institutional rules. Green psychological climate includes organizational and coworker orientations [17], both of which function as proximal environmental cues consistent with social information processing theory; nurses interpret signals from the organization as a whole and from their immediate colleagues to form an overall perception of the workplace's environmental orientation. Sustainable clinical behaviors encompass conservation, waste management, and proactive behaviors, which represent a spectrum from task-related to extra-role green actions. Social learning theory applies across this spectrum: nurses who observe leaders modeling sustainability are more likely to adopt both compliance-oriented behaviors (e.g., proper waste segregation) and discretionary behaviors (e.g., suggesting green initiatives), although the strength of leadership influence may vary across dimensions. Given their theoretical coherence, all three constructs are modeled at the aggregate level in the structural analysis, consistent with established practice in organizational behavior research that treats such constructs as higher-order factors when sub-dimensions share common antecedents and outcomes.

### *2.2. Green Transformational Leadership and Sustainable Clinical Behaviors*

Green transformational leadership represents an extension of transformational leadership theory that specifically focuses on environmental sustainability goals [18]. Green transformational leaders inspire followers through environmental vision, stimulate green creativity, and model sustainable behaviors that employees can observe and replicate [9]. Research has consistently demonstrated that green transformational leadership positively influences employee pro-environmental behaviors across various organizational contexts [9, 19]. Leaders who articulate clear environmental goals and demonstrate personal commitment to sustainability create normative expectations that encourage followers to engage in green behaviors [20].

In healthcare settings, nurse managers serve as immediate supervisors who can significantly influence subordinates' daily practices. According to social learning theory, nurses observe and imitate the environmental behaviors modeled by their managers, particularly when those managers occupy positions of authority and demonstrate consistent commitment to sustainability [15]. This mechanism is particularly salient in clinical settings where nurses face competing demands between patient care and non-clinical tasks. Under conditions of high workload and time pressure, nurses are more likely to rely on immediate supervisory cues as cognitive shortcuts for prioritizing behaviors that fall outside their core clinical responsibilities [7]. Green transformational leaders serve this prioritizing function by signaling that sustainability is not peripheral but integral to the unit's professional expectations. When nurse managers communicate environmental expectations, recognize sustainable practices, and personally demonstrate resource conservation, they establish behavioral norms that nurses are likely to follow. Recent studies confirm that environmentally specific transformational leadership enhances employee environmental awareness and commitment, which subsequently translates into pro-environmental performance [12, 21]. In other words, the influence of green transformational leadership operates not merely through passive observation but through the active reshaping of what nurses perceive as valued and expected within their daily clinical routines. Accordingly, we hypothesize:

**Hypothesis 1**. *Green transformational leadership positively affects sustainable clinical behaviors*.

It is important to note that the existence of formal sustainability protocols in hospitals does not render leadership influence irrelevant. Research consistently demonstrates that organizational environmental policies are necessary but insufficient conditions for employee green behavior; compliance with such policies varies substantially and depends on organizational and leadership factors [22, 23]. In healthcare settings specifically, studies report that 30 to 90% of operating room waste is improperly segregated despite the presence of formal waste management guidelines [4], and that behavioral gains from training interventions erode without sustained organizational reinforcement [8]. Furthermore, sustainable clinical behaviors extend beyond rule-governed actions to include discretionary resource conservation and voluntary proactive initiatives that are not prescribed by hospital protocols. Green transformational leadership is theorized to influence both types of behavior: it enhances compliance with existing green protocols by signaling that sustainability is a genuine organizational priority, and it motivates extra-role environmental actions by inspiring nurses to internalize green values beyond what is formally required [24].

### *2.3. Ethical Climate and Sustainable Clinical Behaviors*

Ethical climate refers to employees' shared perceptions of organizational policies, procedures, and practices that have ethical content [16]. An ethical climate shapes what employees perceive as correct behavior and how ethical issues should be handled within the organization [18]. Research indicates that ethical climate significantly influences various employee attitudes and behaviors, including organizational citizenship behaviors and voluntary initiatives [13, 25]. However, the pathway from ethical climate to environmental behavior requires theoretical elaboration, as ethical climate research has traditionally focused on interpersonal and organizational ethics rather than ecological outcomes. The connection can be understood through the concept of moral extension, whereby the ethical principles embedded in organizational climate broaden employees' scope of moral concern to encompass environmental stakeholders [26]. In nursing specifically, each dimension of ethical climate provides a distinct justification for sustainable behavior: a caring climate extends the professional duty of care to include the wellbeing of communities affected by environmental degradation; a law and code climate aligns sustainability with professional standards, as the ICN Code of Ethics for Nurses explicitly mandates that nurses practice to preserve, sustain, and protect the natural environment [27]; and a rules climate normalizes compliance with hospital green protocols as part of the broader institutional expectation structure.

In healthcare organizations, ethical considerations are fundamental to professional practice. Social information processing theory suggests that nurses interpret the ethical signals embedded in organizational policies and practices to determine what constitutes appropriate behavior [14]. When hospitals maintain a climate that emphasizes caring for stakeholders, adherence to professional codes, and following institutional rules, nurses may perceive environmental stewardship as consistent with their professional ethical obligations. An ethical climate that values the wellbeing of all stakeholders, including future generations affected by environmental degradation, provides moral justification for sustainable practices. Recent evidence supports the positive relationship between organizational ethical climate and employees' voluntary environmental behaviors [13]. Therefore, we propose:

**Hypothesis 2**. *Ethical climate positively affects sustainable clinical behaviors.*

### *2.4. The Mediating Role of Green Psychological Climate*

Green psychological climate refers to employees' perceptions of their organization's and coworkers' environmental orientations, including policies, practices, and values related to sustainability [17]. It represents an individual-level construct that captures how employees interpret the environmental signals within their workplace [28]. According to social information processing theory, employees use such perceptions to guide their behaviors, making green psychological climate a proximal predictor of pro-environmental actions [29]. The rationale for mediation rests on the premise that organizational signals from leaders and ethical climates do not automatically translate into behavioral change. Consistent with the stimulus-organism-response framework [30], organizational stimuli must first be cognitively processed and internalized by employees before they manifest as behavioral responses. Green psychological climate represents this intermediary cognitive state, functioning as the perceptual lens through which nurses interpret whether sustainability is genuinely valued in their workplace and, consequently, whether green behaviors warrant personal investment.

Research demonstrates that green psychological climate serves as a critical mediating mechanism between organizational antecedents and employee green behaviors. Studies have shown that leadership behaviors and organizational practices influence employee pro-environmental behaviors primarily through their effect on green psychological climate [19, 31]. When green transformational leaders communicate environmental values and demonstrate sustainable practices, they shape employees' perceptions that the organization genuinely values environmental responsibility. Similarly, an ethical climate that emphasizes stakeholder welfare and adherence to moral standards contributes to employees' perceptions that sustainability is an organizational priority.

When nurses perceive that their hospital and colleagues genuinely care about environmental issues, they are more likely to internalize these values and translate them into sustainable clinical practices. The green psychological climate serves as a cognitive bridge that connects organizational signals to individual behaviors [32]. Based on this reasoning, we hypothesize:

**Hypothesis 3**. *Green psychological climate mediates the relationship between green transformational leadership and sustainable clinical behaviors.*

**Hypothesis 4**. *Green psychological climate mediates the relationship between ethical climate and sustainable clinical behaviors.*

*2.5. The Moderating Role of Perceived Organizational Hypocrisy*

While positive organizational antecedents can foster sustainable behaviors, their effectiveness may be contingent on employees' perceptions of organizational authenticity. Perceived organizational hypocrisy occurs when employees observe discrepancies between what the organization claims to value and what it actually does [33]. In the context of sustainability, this manifests when employees perceive that their organization or leaders espouse environmental values but fail to act consistently with those values [34].

Research on greenwashing and corporate hypocrisy indicates that such inconsistencies undermine employee trust and reduce engagement in pro-environmental behaviors [35, 11]. When employees perceive greenwashing or hypocrisy, they experience cynicism and moral disengagement, which diminishes their motivation to contribute to organizational sustainability goals [36]. The positive effects of leadership and ethical climate on employee behaviors are contingent on employees believing that organizational commitments are genuine rather than mere rhetoric. The psychological mechanism underlying this moderation can be explained through cognitive dissonance and moral disengagement processes. When nurses observe discrepancies between espoused and enacted sustainability values, they experience cognitive dissonance that they may resolve by disengaging from organizational sustainability efforts [35]. This moral disengagement weakens the otherwise positive link between organizational signals and sustainable behavioral responses.

In healthcare settings, nurses may become skeptical when managers advocate for sustainability but fail to allocate necessary resources, personally model green behaviors, or support sustainable initiatives with tangible actions. Such inconsistencies signal that sustainability is not a genuine organizational priority, potentially neutralizing the positive effects of leadership and ethical climate. Conversely, when perceived hypocrisy is low, the positive signals from green transformational leadership and ethical climate are more likely to translate into sustainable behaviors. Therefore, we propose:

**Hypothesis 5**. Perceived organizational hypocrisy moderates the relationship between green transformational leadership and sustainable clinical behaviors, such that the relationship is weaker when perceived organizational hypocrisy is high.

Notably, the moderating effect of perceived organizational hypocrisy is expected to be particularly pronounced for green transformational leadership compared to ethical climate. Because leadership is a personalized and highly visible influence source, inconsistencies between a specific leader's words and actions are more salient and more directly attributable than inconsistencies in the broader, more diffuse ethical climate [33]. The interpersonal nature of the leader-follower relationship means that perceived hypocrisy carries a stronger sense of betrayal, thereby more forcefully disrupting the behavioral influence of leadership.

**Hypothesis 6**. Perceived organizational hypocrisy moderates the relationship between ethical climate and sustainable clinical behaviors, such that the relationship is weaker when perceived organizational hypocrisy is high.

The proposed conceptual framework is depicted in Figure 1. The model posits that Green Transformational Leadership and Ethical Climate exert direct positive effects on Sustainable Clinical Behaviors (H1 and H2, respectively) as well as indirect effects through Green Psychological Climate, which serves as a mediating mechanism (H3 and H4). Additionally, Perceived Organizational Hypocrisy is hypothesized to function as a boundary condition that weakens the positive direct relationships of both Green Transformational Leadership (H5) and Ethical Climate (H6) with Sustainable Clinical Behaviors. Together, these pathways illustrate how leadership, organizational ethics, and perceptions of authenticity interact to shape nurses' engagement in environmentally sustainable clinical practices.

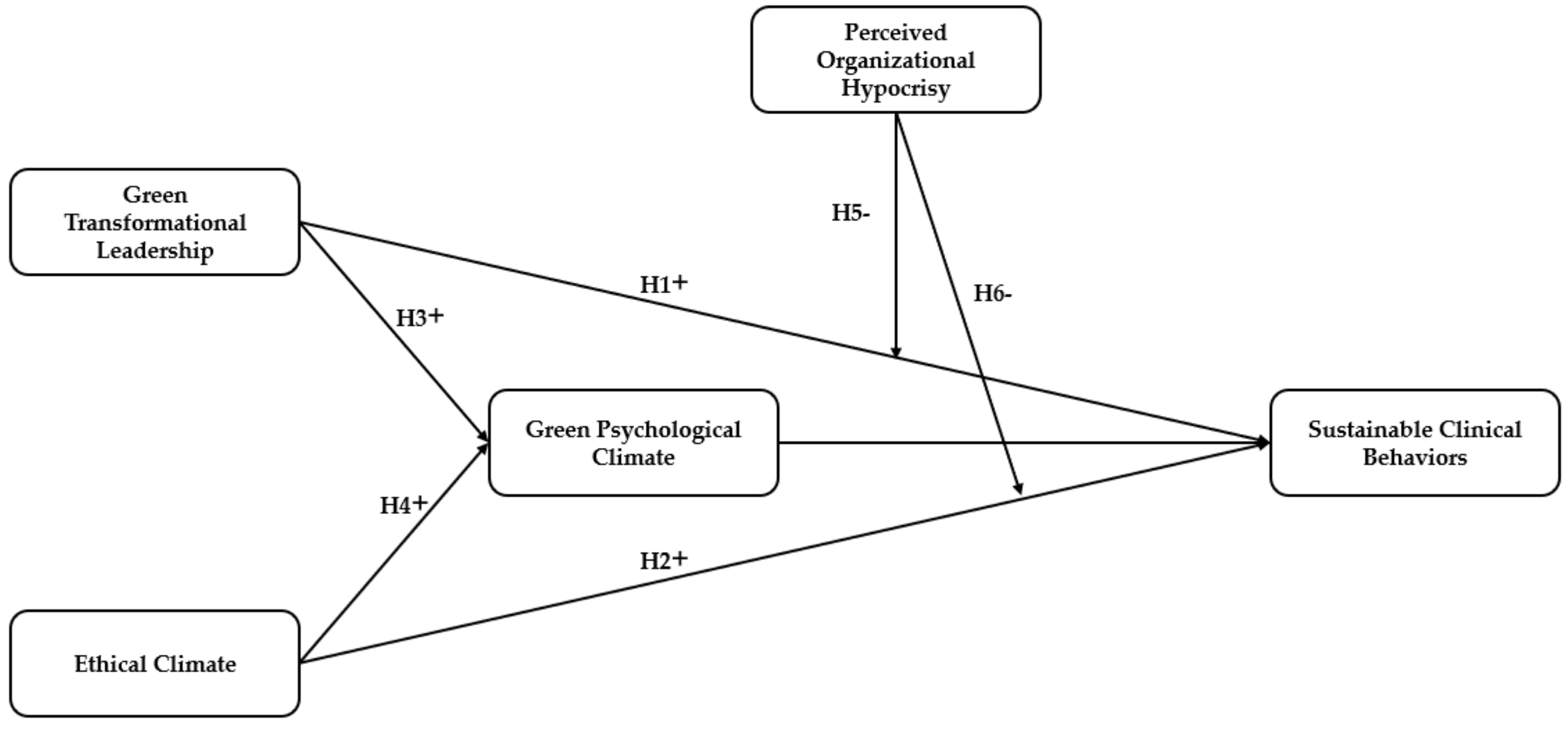

**Figure 1.** Hypothesized conceptual model. (+) indicates a positive hypothesized relationship; (−) indicates a negative moderating effect. H3 and H4 represent mediation through Green Psychological Climate. H5 and H6 represent the moderating role of Perceived Organizational Hypocrisy.

## 3. Methodology

### 3.1. Participants and Procedure

The study targeted registered nurses employed across 11 hospitals in Jordan, including six public and five private institutions located in Amman and other governorates to ensure geographic representation. A stratified random sampling technique was used to recruit participants from various departments such as medical, surgical, intensive care, and emergency units. Inclusion criteria required participants to be full-time registered nurses with at least six months of experience in their current hospital who report directly to a nurse manager. A total of 1,100 questionnaires were distributed during January 2026, coordinated with nursing administration during work shifts. Of these, 760 valid responses were obtained, yielding a response rate of 69.1%. Participation was voluntary and anonymous.

### 3.2. Translation and Pilot Testing

The survey instrument was translated from English to Arabic using a back-translation procedure. Two bilingual experts independently translated the questionnaire into Arabic, and a third expert back-translated it into English. Discrepancies were resolved through discussion to ensure conceptual equivalence. A pilot study was then conducted with 40 nurses from two hospitals not included in the main sample to assess clarity and comprehension. Minor wording adjustments were made based on participant feedback before finalizing the instrument.

### 3.3. Measures

Measurement items were derived from established scales in organizational behavior and environmental psychology, adapted for the Jordanian healthcare context. All items used a five-point Likert scale ranging from 1 (Strongly Disagree) to 5 (Strongly Agree) unless otherwise noted. The complete list of measurement items and their sources is presented in Appendix Table A1.

Green Transformational Leadership was assessed using six items adapted from Chen and Chang [18]. Ethical Climate was measured using 12 items from Victor and Cullen [16], comprising three subscales: Caring (5 items), Law and Code (4 items), and Rules (3 items). Green Psychological Climate was evaluated using eight items from Norton et al. [17], capturing organizational (4 items) and coworker (4 items) environmental orientations. Perceived Organizational Hypocrisy was measured using four items adapted from Dineen et al. [37], modified to focus on sustainability. The items for perceived organizational hypocrisy were operationalized at the supervisor level because, for bedside nurses, the nurse manager is the most immediate and salient organizational representative. Research on social information processing theory suggests that employees primarily interpret organizational signals through their direct supervisors, who serve as proximal embodiments of organizational values and priorities [14]. This approach is consistent with prior leadership research in which supervisor-level measurement is used as a proxy for employees' experience of broader organizational phenomena [33].

Sustainable Clinical Behaviors were assessed using a 13-item scale developed for this study based on Robertson and Barling [34], encompassing three dimensions: Conservation (5 items), Waste Management (4 items), and Proactive Behaviors (4 items). Conservation items reflect primarily discretionary resource-saving decisions within clinical routines, Waste Management items include both protocol-adherent and discretionary supply management behaviors, and Proactive Behaviors capture entirely voluntary, extra-role environmental initiatives. This multidimensional conceptualization aligns with the established distinction in the employee green behavior literature between required (task-related) and voluntary (extra-role) pro-environmental actions [22]. This variable used frequency-based responses ranging from 1 (Never) to 5 (Always).

Three control variables were included in the structural model: gender, age, and professional experience. Gender was measured as a binary variable (0 = female, 1 = male). These variables were controlled based on prior empirical evidence. Age has been associated with stronger pro-environmental attitudes and behaviors, as older individuals tend to exhibit greater environmental concern and responsibility [38]. Professional experience may influence the extent to which nurses have internalized organizational norms and routines related to sustainability [39]. Gender has been consistently linked to pro-environmental behavior, with research suggesting that women tend to report higher levels of environmental concern and engagement in sustainable practices [40].

Education level and hospital type were collected as demographic variables but were not included as control variables in the structural model. Preliminary analyses revealed that neither education level ($F(2, 757) = 0.24$, $p = 0.785$) nor hospital type ($t(742.78) = 0.92$, $p = 0.358$) were significantly associated with sustainable clinical behaviors. Including non-significant control variables can reduce statistical power and introduce unnecessary complexity into the model without improving estimation accuracy [41, 42]. Furthermore, the inclusion of irrelevant control variables in structural equation models has been shown to potentially bias parameter estimates and inflate Type I or Type II error rates [42]. Therefore, consistent with best-practice recommendations for the judicious use of control variables, education level and hospital type were excluded from the structural model.

### *3.4. Data Analysis*

Data were analyzed using structural equation modeling with maximum likelihood estimation. The measurement model was first evaluated through confirmatory factor analysis to establish reliability and validity. The structural model was then tested to examine direct effects, mediation, and moderation hypotheses. Bootstrapping with 5,000 subsamples was employed to generate confidence intervals for indirect and interaction effects.

## 4. Results

### *4.1. Demographic Characteristics and Descriptive Statistics*

The study analyzed data collected from a total of 760 registered nurses working in public and private hospitals in Jordan. Table 1 presents the demographic characteristics of the sample. The demographic profile indicated a diverse sample, with 39.3% of respondents being female and 60.7% male. Although nursing is predominantly female globally, with approximately 90% of the world's nurses being women [43], the Jordanian context is notably different. Male representation in Jordanian nursing is substantially higher than global averages, a pattern driven by economic factors, the expansion of healthcare facilities, and the historical role of military nursing in attracting men to the profession [44]. According to the Jordanian Nursing and Midwifery Council, men account for approximately 30% of the total nursing workforce [45]. However, male representation has been rising over time and varies considerably across settings and cohorts. Data from the Jordanian Nursing Council indicate that the ratio of male to female nursing students reached 60/40 by 2009 [46, 47], and Saleh et al. [48] reported that the proportion of males among registered nurses increased from 38% in 2009 to as high as 65% in certain institutional registries by 2012. These higher figures likely reflect the composition of more recent graduate cohorts rather than the overall workforce, and the proportion of males tends to be higher in public and military hospitals than in private institutions [47]. The gender composition of the present sample (60.7% male) is therefore higher than the national workforce average but is consistent with the demographic profile of the hospital-based Jordanian nursing workforce from which the sample was drawn, particularly given the inclusion of six public hospitals in the sampling frame. To ensure transparency, we note that gender was controlled for in the structural model, and it was not a significant predictor of sustainable clinical behaviors (B = 0.031, p = 0.201), suggesting that the gender composition does not substantively influence the study's core findings.

The participants ranged in age from 22 to 65 years, with a mean age of 34.21 years (SD = 7.73). Regarding professional experience, the sample had an average tenure of 12.39 years (SD = 7.74), ensuring that respondents possessed sufficient organizational knowledge to assess leadership and climate factors accurately. The sample was fairly balanced between hospital types, with 53.4% working in public hospitals and 46.6% in private institutions. In terms of educational attainment, 18.3% held a diploma, 64.5% held a bachelor's degree, and 17.2% held a master's degree or higher.

**Table 1.** Demographic Characteristics of Participants (N = 760)

| Variable | Category | Frequency | Percentage |
|---|---|---|---|
| Gender | Female | 299 | 39.3% |
| | Male | 461 | 60.7% |
| Hospital Type | Public | 406 | 53.4% |
| | Private | 354 | 46.6% |
| Education | Diploma | 139 | 18.3% |
| | Bachelor | 490 | 64.5% |
| | Master or higher | 131 | 17.2% |
| Continuous Variables | Mean | SD | Range |
| Age (years) | 34.21 | 7.73 | 22-65 |
| Experience (years) | 12.39 | 7.74 | 0.5-35 |

Descriptive statistics for the study constructs are presented in Table 2. Participants generally perceived moderate to high levels of the measured variables. Sustainable Clinical Behaviors had the highest mean score (M = 3.69, SD = 0.73), followed closely by Ethical Climate (M = 3.59, SD = 0.67) and Perceived Organizational Hypocrisy (M = 3.58, SD = 0.79). Green Transformational Leadership (M = 3.58, SD = 0.78) and Green Psychological Climate (M = 3.57, SD = 0.75) also demonstrated moderate levels. The standard deviations suggested adequate variability in responses, and the skewness and kurtosis values fell within the acceptable range of plus or minus two, indicating approximate normality of the data distribution.

**Table 2.** Descriptive Statistics for Study Constructs (N = 760)

| Construct | Mean | SD | Min | Max | Skewness | Kurtosis |
|---|---|---|---|---|---|---|
| Green Transformational Leadership | 3.58 | 0.78 | 1.50 | 5.00 | -0.21 | -0.44 |
| Ethical Climate | 3.59 | 0.67 | 1.58 | 5.00 | -0.25 | -0.27 |
| Green Psychological Climate | 3.57 | 0.75 | 1.13 | 5.00 | -0.33 | -0.18 |
| Perceived Organizational Hypocrisy | 3.58 | 0.79 | 1.00 | 5.00 | -0.21 | -0.36 |
| Sustainable Clinical Behaviors | 3.69 | 0.73 | 1.38 | 5.00 | -0.38 | -0.23 |

*4.2. Measurement Model Evaluation*

To establish the appropriateness of the proposed factor structure, a series of confirmatory factor analyses were conducted comparing the hypothesized ten-factor measurement model against four alternative models (Table 3). Model 1, a single-factor model in which all items loaded on one latent factor, exhibited poor fit (CFI = 0.408, RMSEA = 0.150). Model 2, a two-factor model distinguishing predictor variables from outcome variables, showed marginal improvement but remained inadequate (CFI = 0.521, RMSEA = 0.135). Model 3, a three-factor model grouping items into leadership/ethics, climate/hypocrisy, and behavioral dimensions, also failed to achieve acceptable fit (CFI = 0.613, RMSEA = 0.122). Model 4, a five-factor model representing the five main constructs without distinguishing sub-dimensions, improved considerably but still fell short of recommended thresholds (CFI = 0.813, RMSEA = 0.085). The hypothesized ten-factor model demonstrated a substantially superior fit ($\chi^2$ = 827.83, df = 815, p = 0.370; CFI = 0.999, TLI = 0.999, RMSEA = 0.005, SRMR = 0.020), confirming that the proposed measurement structure best represents the data and that the constructs are empirically distinct.

**Table 3.** Confirmatory Factor Analysis: Model Comparison

| Model | $\chi^2$ | df | p | CFI | TLI | RMSEA | SRMR |
|---|---|---|---|---|---|---|---|
| Model 1: Single-factor | 15641.75 | 860 | < 0.001 | 0.408 | 0.379 | 0.150 | 0.143 |
| Model 2: Two-factor | 12814.31 | 859 | < 0.001 | 0.521 | 0.497 | 0.135 | 0.125 |
| Model 3: Three-factor | 10522.57 | 857 | < 0.001 | 0.613 | 0.592 | 0.122 | 0.119 |
| Model 4: Five-factor | 5522.07 | 850 | < 0.001 | 0.813 | 0.801 | 0.085 | 0.053 |
| Model 5: Hypothesized 10-factor | 827.83 | 815 | 0.370 | 0.999 | 0.999 | 0.005 | 0.020 |

*Note.* Model 1 = all items on a single factor; Model 2 = all predictor items vs. all outcome items; Model 3 = leadership/ethics, climate/hypocrisy, and behaviors; Model 4 = five main constructs (GTL, EC, GPC, POH, SCB) without sub-dimensions; Model 5 = hypothesized ten-factor measurement model.

The measurement model was assessed to establish reliability and validity before testing the structural relationships. Factor loadings for all items were examined to ensure indicator reliability. As presented in Table 4, all standardized factor loadings exceeded the recommended threshold of 0.70, ranging from 0.81 to 0.90, with all values demonstrating strong relationships with their respective constructs. This confirms that the items share a high proportion of variance with their respective constructs. Internal consistency reliability was assessed using Cronbach's alpha and Composite Reliability. All constructs demonstrated excellent reliability, with Cronbach's alpha values ranging from 0.873 to 0.936. Composite Reliability values were well above the 0.70 threshold, ranging from 0.865 to 0.937. Convergent validity was established as the Average Variance Extracted for all constructs exceeded the 0.50 benchmark, with values ranging from 0.506 to 0.711, indicating that the constructs explain more than half of the variance of their indicators.

**Table 4.** Psychometric Properties of the Measurement Model

| Construct / Items | Factor Loading | Cronbach's $\alpha$ | CR | AVE | VIF |
|---|---|---|---|---|---|
| Green Transformational Leadership (GTL) | | 0.936 | 0.937 | 0.711 | 1.35 |
| GTL1 | 0.838 | | | | 2.879 |
| GTL2 | 0.866 | | | | 3.302 |
| GTL3 | 0.862 | | | | 3.182 |
| GTL4 | 0.823 | | | | 2.694 |
| GTL5 | 0.847 | | | | 3.008 |
| GTL6 | 0.824 | | | | 2.724 |
| Ethical Climate - Caring | | 0.917 | 0.917 | 0.689 | 1.77 |
| EC_Caring1 | 0.810 | | | | 2.448 |
| EC_Caring2 | 0.836 | | | | 2.683 |
| EC_Caring3 | 0.847 | | | | 2.843 |
| EC_Caring4 | 0.844 | | | | 2.793 |
| EC_Caring5 | 0.811 | | | | 2.468 |
| Ethical Climate - Law and Code | | 0.907 | 0.907 | 0.710 | 1.75 |

| EC_Law1 | 0.841 | | | | 2.628 |
|---|---|---|---|---|---|
| EC_Law2 | 0.853 | | | | 2.769 |
| EC_Law3 | 0.816 | | | | 2.504 |
| EC_Law4 | 0.859 | | | | 2.824 |
| Ethical Climate – Rules | | 0.873 | 0.874 | 0.699 | 1.76 |
| EC_Rules1 | 0.827 | | | | 2.315 |
| EC_Rules2 | 0.845 | | | | 2.391 |
| EC_Rules3 | 0.836 | | | | 2.352 |
| Green Psychological Climate – Organization | | 0.903 | 0.903 | 0.700 | 1.79 |
| GPC_Org1 | 0.838 | | | | 2.629 |
| GPC_Org2 | 0.819 | | | | 2.448 |
| GPC_Org3 | 0.833 | | | | 2.509 |
| GPC_Org4 | 0.858 | | | | 2.821 |
| Green Psychological Climate – Coworker | | 0.896 | 0.897 | 0.684 | 1.86 |
| GPC_Coworker1 | 0.814 | | | | 2.314 |
| GPC_Coworker2 | 0.819 | | | | 2.448 |
| GPC_Coworker3 | 0.840 | | | | 2.568 |
| GPC_Coworker4 | 0.836 | | | | 2.540 |
| Perceived Organizational Hypocrisy (POH) | | 0.908 | 0.908 | 0.711 | 1.22 |
| POH1 | 0.841 | | | | 2.647 |
| POH2 | 0.858 | | | | 2.834 |
| POH3 | 0.844 | | | | 2.692 |
| POH4 | 0.829 | | | | 2.547 |
| Sustainable Clinical Behaviors – Conservation | | 0.927 | 0.927 | 0.718 | |
| SCB_Cons1 | 0.848 | | | | |
| SCB_Cons2 | 0.846 | | | | |
| SCB_Cons3 | 0.852 | | | | |
| SCB_Cons4 | 0.834 | | | | |
| SCB_Cons5 | 0.855 | | | | |
| Sustainable Clinical Behaviors – Waste Management | | 0.936 | 0.937 | 0.787 | |
| SCB_Waste1 | 0.903 | | | | |
| SCB_Waste2 | 0.872 | | | | |
| SCB_Waste3 | 0.891 | | | | |
| SCB_Waste4 | 0.881 | | | | |
| Sustainable Clinical Behaviors – Proactive | | 0.909 | 0.910 | 0.716 | |
| SCB_Proactive1 | 0.831 | | | | |
| SCB_Proactive2 | 0.843 | | | | |
| SCB_Proactive3 | 0.847 | | | | |
| SCB_Proactive4 | 0.862 | | | | |

Note. CR = Composite Reliability; AVE = Average Variance Extracted. All factor loadings are significant at $p < 0.001$.

To assess multicollinearity, Variance Inflation Factor values were computed at both the construct and item levels, as reported in Table 4. At the construct level, VIF values ranged from 1.22 (Perceived Organizational Hypocrisy) to 1.86 (Green Psychological Climate – Coworker), all well below the threshold of 3.0. At the item level, VIF values ranged from 2.31 to 3.30, with all values remaining below the conservative threshold of 5.0 [49]. The slightly higher item-level VIF values observed for some Green Transformational Leadership indicators (e.g., GTL2 = 3.30) are consistent with the high internal consistency of this construct ($\alpha$ = 0.936) and do not indicate problematic multicollinearity. These results confirm that multicollinearity was not a concern for either the measurement or structural analyses.

The confirmatory factor analysis revealed an excellent fit to the data. The chi-square value was 827.83 with 815 degrees of freedom, yielding a non-significant p-value of 0.370, which indicates that the proposed model adequately reproduces the observed covariance matrix. The Comparative Fit Index was 0.999 and the Tucker-Lewis Index was 0.999, both substantially exceeding the recommended threshold of 0.95. The Root Mean Square Error of Approximation was 0.005, well below the stringent criterion of 0.06, and the Standardized Root Mean Square Residual was 0.020, comfortably below the 0.08 threshold. These indices collectively demonstrate an exceptional fit between the hypothesized measurement model and the observed data.

Discriminant validity was assessed using both the Fornell-Larcker criterion and the Heterotrait-Monotrait ratio of correlations. Table 5 presents the results of Fornell-Larcker Criterion. The square root of the AVE for each construct (diagonal values in bold) was greater than its highest correlation with any other construct, satisfying the Fornell-Larcker criterion. Furthermore, all HTMT values were well below the strict threshold of 0.85, with the highest observed value being 0.58 between Green Psychological Climate and Sustainable Clinical Behaviors. This confirms that the constructs are empirically distinct from one another.

**Table 5.** Discriminant Validity: Fornell-Larcker Criterion

| Construct | GTL | EC | GPC | POH | SCB |
|---|---|---|---|---|---|
| 1. Green Transformational Leadership | **0.843** | | | | |
| 2. Ethical Climate | 0.410 | **0.711** | | | |
| 3. Green Psychological Climate | 0.396 | 0.350 | **0.751** | | |
| 4. Perceived Organizational Hypocrisy | −0.302 | −0.393 | −0.194 | **0.843** | |
| 5. Sustainable Clinical Behaviors | 0.449 | 0.395 | 0.551 | −0.332 | **0.737** |

Note. Bold diagonal values represent the square root of the Average Variance Extracted (AVE). Off-diagonal values are Pearson correlations between constructs. Discriminant validity is established when each diagonal value exceeds all correlations in its row and column.

To assess the potential issue of Common Method Bias, Harman's single-factor test was conducted. The single-factor model yielded a poor fit with a CFI of 0.408, RMSEA of 0.150, and SRMR of 0.143, indicating that a single factor did not account for the majority of the variance and suggesting that CMB was not a pervasive issue in this study.

### *4.3. Structural Model and Hypothesis Testing*

The structural model was evaluated to test the hypothesized relationships among the constructs. The model fit indices for the structural model demonstrate acceptable fit with CFI of 0.996, TLI of 0.995, RMSEA of 0.013, and SRMR of 0.061. The model demonstrated good explanatory power, with the predictors explaining approximately 35.7% of the variance in Sustainable Clinical Behaviors ($R^2 = 0.357$) and 16.8% of the variance in Green Psychological Climate ($R^2 = 0.168$).

The effect sizes were evaluated using partial eta-squared values, as shown in Table 6. Green Transformational Leadership demonstrated a large effect size (partial $\eta^2 = 0.215$), Green Psychological Climate showed a large effect (partial $\eta^2 = 0.156$), Ethical Climate exhibited a medium effect (partial $\eta^2 = 0.075$), and Perceived Organizational Hypocrisy had a small effect (partial $\eta^2 = 0.023$). These effect sizes indicate meaningful practical significance for the key predictors in the model.

**Table 6.** Effect Sizes (Partial Eta-Squared) for Predictors of Sustainable Clinical Behaviors

| Predictor | Partial $\eta^2$ | 95% CI Lower | Effect Size Interpretation |
|---|---|---|---|
| Green Transformational Leadership | 0.215 | 0.174 | Large |
| Ethical Climate | 0.075 | 0.048 | Medium |
| Green Psychological Climate | 0.156 | 0.118 | Large |
| Perceived Organizational Hypocrisy | 0.023 | 0.008 | Small |

Note. Effect size interpretation: small = 0.01, medium = 0.06, large = 0.14 (Cohen, 1988).

The results of the structural model and hypothesis testing are presented in Table 7. The direct relationships were tested using bootstrapping with 5000 subsamples. Green Transformational Leadership exerted a significant positive direct effect on Sustainable Clinical Behaviors (β = 0.215, t = 6.97, p < 0.001), supporting the hypothesis that green leadership directly promotes sustainable behaviors among nurses. Similarly, Ethical Climate demonstrated a significant positive effect on Sustainable Clinical Behaviors (β = 0.161, t = 5.06, p < 0.001), confirming that an ethical organizational environment facilitates pro-environmental clinical practices. Green Transformational Leadership also significantly predicted Green Psychological Climate (β = 0.373, t = 13.53, p < 0.001), as did Ethical Climate (β = 0.453, t = 16.94, p < 0.001). Green Psychological Climate, in turn, significantly influenced Sustainable Clinical Behaviors (β = 0.427, t = 13.74, p < 0.001). Perceived Organizational Hypocrisy demonstrated a significant negative direct effect on Sustainable Clinical Behaviors (β = -0.109, t = -4.16, p < 0.001), indicating that when nurses perceive inconsistency between organizational rhetoric and actions regarding sustainability, their engagement in sustainable behaviors diminishes.

**Table 7.** Structural Model Results and Hypothesis Testing

| Hypothesis | Path | β | SE | t-value | p-value | 95% CI | Decision |
|---|---|---|---|---|---|---|---|
| | | | | Direct Effects | | | |
| H1 | GTL → SCB | 0.215 | 0.031 | 6.97 | < 0.001 | [0.150, 0.272] | Supported |
| H2 | EC → SCB | 0.161 | 0.032 | 5.06 | < 0.001 | [0.099, 0.220] | Supported |
| | GTL → GPC | 0.373 | 0.028 | 13.53 | < 0.001 | [0.320, 0.425] | - |
| | EC → GPC | 0.453 | 0.027 | 16.94 | < 0.001 | [0.401, 0.507] | - |
| | GPC → SCB | 0.427 | 0.031 | 13.74 | < 0.001 | [0.367, 0.488] | - |
| | POH → SCB | -0.109 | 0.026 | -4.16 | < 0.001 | [-0.160, -0.058] | - |
| | | | | Indirect Effects | | | |
| H3 | GTL → GPC → SCB | 0.122 | 0.019 | 6.44 | < 0.001 | [0.086, 0.162] | Supported |
| H4 | EC → GPC → SCB | 0.092 | 0.020 | 4.63 | < 0.001 | [0.054, 0.131] | Supported |
| | | | | Moderation Effects | | | |
| H5 | GTL × POH → SCB | -0.153 | 0.026 | -5.98 | < 0.001 | [-0.195, -0.103] | Supported |
| H6 | EC × POH → SCB | -0.065 | 0.025 | -2.61 | 0.009 | [-0.118, -0.017] | Supported |
| | | | | Control Variables | | | |
| | Age → SCB | 0.240 | 0.100 | 2.39 | 0.017 | [0.044, 0.423] | Significant |
| | Experience → SCB | -0.258 | 0.101 | -2.55 | 0.011 | [-0.440, -0.056] | Significant |
| | Gender → SCB | 0.031 | 0.024 | 1.28 | 0.201 | [-0.017, 0.079] | Not Significant |

Note. GTL = Green Transformational Leadership; EC = Ethical Climate; GPC = Green Psychological Climate; POH = Perceived Organizational Hypocrisy; SCB = Sustainable Clinical Behaviors. Bootstrap samples = 5000. β = standardized path coefficient; SE = standard error; CI = confidence interval.

Regarding the control variables, age was found to positively influence sustainable behaviors (β = 0.240, t = 2.39, p < 0.05), while experience showed a negative relationship (β = -0.258, t = -2.55, p < 0.05), likely due to suppression effects given the high correlation between age and experience. Gender did not have a significant effect on the outcome ($\beta$ = 0.031, t = 1.28, p = 0.201).

*4.4. Mediation Analysis*

Mediation analysis was conducted to test the indirect effects of Green Transformational Leadership and Ethical Climate on Sustainable Clinical Behaviors through Green Psychological Climate, utilizing bootstrapping with 5000 subsamples to generate confidence intervals for the indirect effects. As shown in Table 7, the results confirmed that Green Psychological Climate significantly mediates the relationship between Green Transformational Leadership and Sustainable Clinical Behaviors. The indirect effect was 0.122 ($p < 0.001$), with a 95% confidence interval of [0.086, 0.162] that did not include zero, providing strong evidence for mediation and supporting H3. The total effect of Green Transformational Leadership on Sustainable Clinical Behaviors was 0.319, comprising both the direct effect (0.215) and the indirect effect through Green Psychological Climate (0.122). Furthermore, Green Psychological Climate was found to significantly mediate the relationship between Ethical Climate and Sustainable Clinical Behaviors, supporting H4. The indirect effect was 0.092 ($p < 0.001$), with a 95% confidence interval of [0.054, 0.131] excluding zero. The total effect of Ethical Climate on Sustainable Clinical Behaviors was 0.205. These findings suggest that leadership and ethical signals within the hospital promote sustainable behaviors partially by fostering a supportive psychological climate regarding environmental issues.

Both relationships exhibited complementary partial mediation, as the direct and indirect effects were statistically significant and operated in the same positive direction. This indicates that Green Psychological Climate partially, but not fully, accounts for the effects of Green Transformational Leadership and Ethical Climate on Sustainable Clinical Behaviors, suggesting that additional mediating mechanisms may also be at work.

*4.5. Moderation Analysis*

To test the moderating role of Perceived Organizational Hypocrisy, interaction terms were introduced into the structural model. As presented in Table 7, the interaction term between Green Transformational Leadership and Perceived Organizational Hypocrisy was negative and statistically significant ($\beta$ = -0.153, t = -5.98, $p < 0.001$), supporting H5. This finding indicates that the positive impact of green leadership on sustainable behaviors is weakened when nurses perceive high levels of organizational hypocrisy. To visualize this interaction, a simple slope analysis was conducted and the results are displayed in Figure 2. The plot reveals that the relationship between Green Transformational Leadership and Sustainable Clinical Behaviors is steeper and more positive when hypocrisy is low (one standard deviation below the mean), whereas the relationship becomes flatter and weaker when hypocrisy is high (one standard deviation above the mean). This pattern confirms that perceived hypocrisy acts as a boundary condition that attenuates the effectiveness of green transformational leadership.

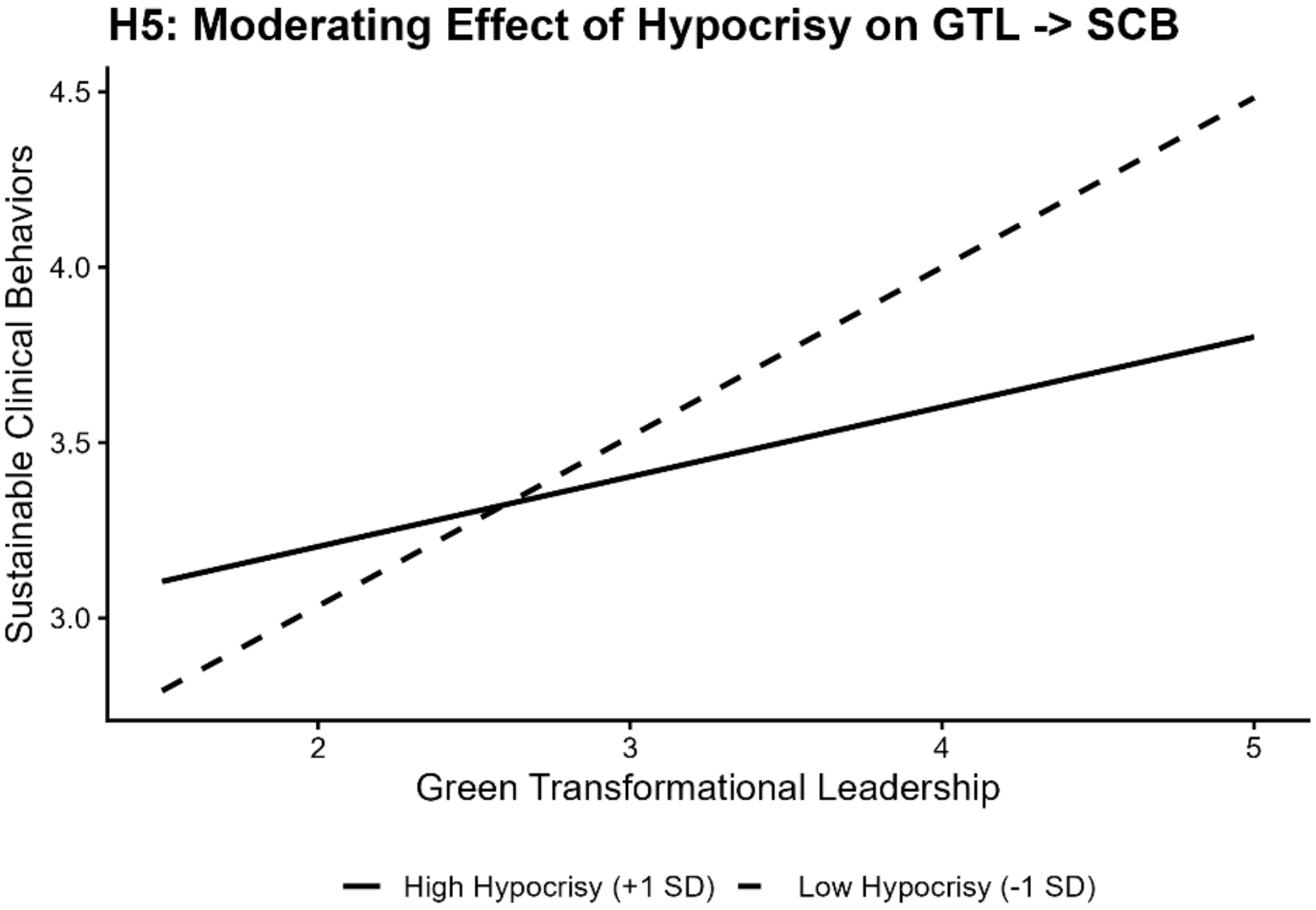

**Figure 2.** Interaction Effect of Perceived Organizational Hypocrisy on the Relationship between Green Transformational Leadership and Sustainable Clinical Behaviors.The solid line represents the relationship at high levels of perceived organizational hypocrisy (+1 SD). The dashed line represents the relationship at low levels of perceived organizational hypocrisy (-1 SD).

Similarly, the interaction between Ethical Climate and Perceived Organizational Hypocrisy was found to be negative and significant ($\beta = -0.065$, $t = -2.61$, $p = 0.009$), supporting H6. This finding suggests that Perceived Organizational Hypocrisy also acts as a boundary condition for ethical climate. The positive influence of an ethical climate on sustainable clinical behaviors is diminished in the presence of perceived hypocrisy. Figure 3 illustrates this interaction, showing that the positive slope of Ethical Climate on the outcome is attenuated for individuals who perceive higher levels of hypocrisy within the organization.

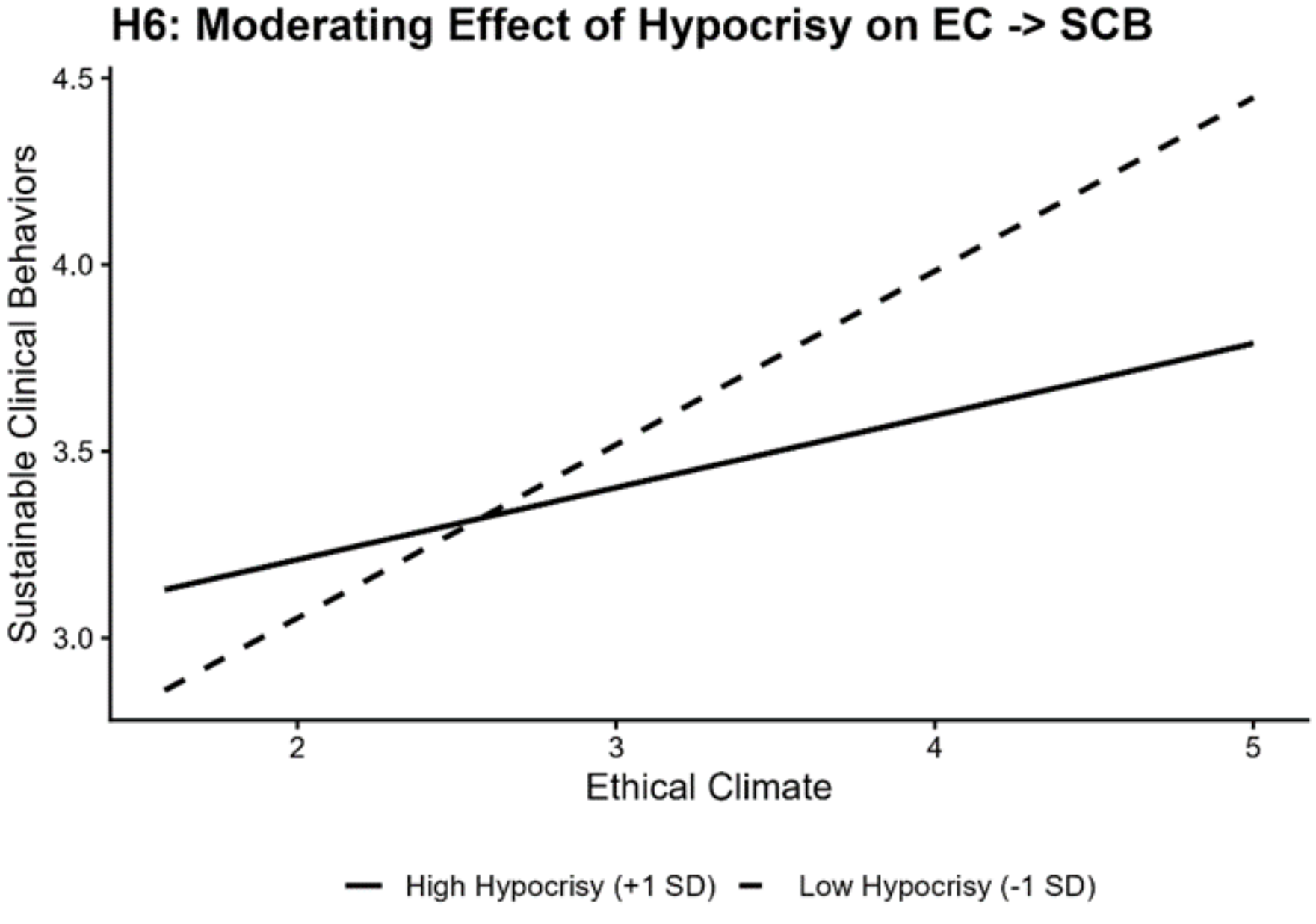

**Figure 3.** Interaction Effect of Perceived Organizational Hypocrisy on the Relationship between Ethical Climate and Sustainable Clinical Behaviors.The solid line represents the relationship at high

levels of perceived organizational hypocrisy (+1 SD). The dashed line represents the relationship at low levels of perceived organizational hypocrisy (-1 SD).

When nurses observe that the hospital espouses environmental values but fails to act consistently with those values, the beneficial effects of both green leadership and ethical climate on their sustainable behaviors are undermined. These moderation findings highlight the critical importance of organizational authenticity in sustainability initiatives. Leaders must not only communicate green values but also demonstrate genuine commitment through consistent actions and resource allocation to maximize the effectiveness of their sustainability efforts. Table 8 provides a summary of all hypothesis testing results.

**Table 8.** Summary of Hypothesis Testing Results

| Hypothesis | Description | Result |
|---|---|---|
| H1 | Green Transformational Leadership has a positive direct effect on Sustainable Clinical Behaviors | Supported |
| H2 | Ethical Climate has a positive direct effect on Sustainable Clinical Behaviors | Supported |
| H3 | Green Psychological Climate mediates the relationship between Green Transformational Leadership and Sustainable Clinical Behaviors | Supported |
| H4 | Green Psychological Climate mediates the relationship between Ethical Climate and Sustainable Clinical Behaviors | Supported |
| H5 | Perceived Organizational Hypocrisy moderates the relationship between Green Transformational Leadership and Sustainable Clinical Behaviors, such that the relationship is weaker when hypocrisy is high | Supported |
| H6 | Perceived Organizational Hypocrisy moderates the relationship between Ethical Climate and Sustainable Clinical Behaviors, such that the relationship is weaker when hypocrisy is high | Supported |

## 5. Discussion

This study examined the organizational determinants of sustainable clinical behaviors among registered nurses in Jordanian hospitals, focusing on the roles of green transformational leadership, ethical climate, green psychological climate, and perceived organizational hypocrisy. All six hypotheses were supported, providing a comprehensive understanding of how organizational factors promote or inhibit nurses' engagement in environmentally sustainable practices.

### *5.1. Main Findings*

The results demonstrated that green transformational leadership positively and significantly influenced sustainable clinical behaviors among nurses ($\beta = 0.215$, $p < 0.001$), supporting H1. This finding indicates that nurse managers who articulate environmental visions, model sustainable practices, and stimulate staff to think about green ideas can effectively promote resource conservation, waste management, and proactive environmental behaviors among their subordinates. This result aligns with recent meta-analytic evidence showing that environmentally specific transformational leadership consistently predicts employee green behaviors across organizational contexts [9]. The effect size was large (partial $\eta^2 = 0.215$), suggesting that green transformational leadership represents a practically significant lever for enhancing sustainability in clinical settings.

Similarly, ethical climate demonstrated a positive effect on sustainable clinical behaviors ($\beta = 0.161$, $p < 0.001$), confirming H2. This finding suggests that when hospitals maintain climates characterized by caring for stakeholders, adherence to professional codes, and following institutional rules, nurses are more inclined to engage in environmentally responsible practices. The relationship between ethical climate and sustainable behaviors can be understood through the lens of moral extension, whereby nurses who operate in ethically oriented environments may perceive environmental stewardship as consistent with their professional duty of care. This finding extends prior research on ethical climate and organizational citizenship behaviors [18] to the domain of environmental sustainability.

The mediation hypotheses (H3 and H4) were both supported, revealing that green psychological climate serves as a significant mechanism through which leadership and ethical climate influence sustainable behaviors. The indirect effect of green transformational leadership through green psychological climate was 0.122 ($p < 0.001$), while the indirect effect of ethical climate was 0.092 ($p < 0.001$). These findings indicate that organizational signals about sustainability are processed through employees' perceptions of the workplace green climate before translating into behavioral outcomes. This is consistent with social information processing theory, which posits that employees rely on environmental cues to guide their attitudes and behaviors [14]. Recent studies have similarly demonstrated the mediating role of green psychological climate in linking organizational antecedents to employee green behaviors [19, 28].

Notably, the moderation hypotheses (H5 and H6) were confirmed, revealing that perceived organizational hypocrisy weakens the positive relationships between both green transformational leadership ($\beta = -0.153$, $p < 0.001$) and ethical climate ($\beta = -0.065$, $p < 0.01$) with sustainable clinical behaviors. These findings highlight that the effectiveness of positive organizational factors is contingent on employees perceiving consistency between organizational rhetoric and action. When nurses observe that their managers advocate for sustainability but fail to demonstrate personal commitment or allocate necessary resources, the motivating effects of leadership and ethical climate are attenuated. This result aligns with emerging research on greenwashing and corporate hypocrisy, which demonstrates that perceived inconsistencies undermine employee trust and reduce engagement in pro-environmental behaviors 35, 36].

The simple slope analyses illustrated these interaction effects visually. Under conditions of low perceived hypocrisy, the relationship between green transformational leadership and sustainable clinical behaviors was steep and positive, whereas under high hypocrisy conditions, the relationship became substantially flatter. Similar patterns emerged for ethical climate, though the moderating effect was smaller in magnitude. These findings underscore the critical importance of authenticity in organizational sustainability initiatives.

### *5.2. Theoretical Contributions*

This study makes several theoretical contributions to the literature. First, it extends the application of green transformational leadership theory to the healthcare sector, an underexplored context despite the industry's substantial environmental footprint. While prior research has predominantly examined green leadership in manufacturing and hospitality settings [9, 21], this study demonstrates that the principles of environmentally specific leadership are equally applicable to clinical environments where nurses must balance patient care demands with sustainability considerations.

Second, the study integrates ethical climate research with environmental sustainability literature. Previous studies have examined ethical climate primarily in relation to organizational citizenship behaviors, whistleblowing, and counterproductive work behaviors [25, 50]. By demonstrating that ethical climate also predicts sustainable behaviors, this research expands the nomological network of ethical climate and responds to calls for examining how organizational ethics relate to pro-environmental outcomes [13].

Third, this study advances understanding of the boundary conditions affecting sustainability initiatives by introducing perceived organizational hypocrisy as a moderator. While the positive antecedents of employee green behavior have been extensively documented, the factors that may undermine these relationships have received comparatively less attention [11]. By demonstrating that perceived hypocrisy attenuates the effects of both leadership and ethical climate, this research highlights the "dark side" of organizational perceptions in sustainability contexts and contributes to the emerging literature on greenwashing effects on employees [35].

Fourth, the study contributes to contextualized understanding of organizational behavior by providing evidence from Jordan, a developing nation in the Middle East facing unique resource constraints. Research on healthcare sustainability has been predominantly conducted in Western contexts, limiting the generalizability of findings to regions with different cultural, economic, and institutional characteristics. Jordan's position as a resource-scarce country where water scarcity and energy costs create both economic and ecological imperatives for conservation provides a valuable context for understanding sustainability behaviors under conditions of genuine resource constraint.

Beyond these context-specific contributions, the findings speak to broader debates in the transformational leadership and organizational sustainability literatures. First, the results reinforce the general principle that transformational leadership operates through shaping followers' perceptions of their work environment before influencing behavior [51], as evidenced by the mediating role of green psychological climate. This suggests that the climate-shaping mechanism identified in general transformational leadership research extends to the environmental domain and holds even in highly regulated professional contexts where formal protocols exist. Second, by demonstrating that leadership and ethical climate jointly predict sustainable behaviors encompassing both required and voluntary actions, this study contributes to the broader employee green behavior literature, which has called for research examining how organizational factors differentially influence task-related versus proactive pro-environmental behaviors [22, 24]. The finding that green transformational leadership promotes sustainable clinical behaviors across this behavioral spectrum suggests that environmentally specific leadership may be particularly effective because it simultaneously signals organizational expectations (enhancing compliance) and inspires personal commitment (motivating extra-role action). Collectively, these findings suggest that the theoretical mechanisms identified in this study are not unique to healthcare but may generalize to other professional contexts where employee sustainability behaviors operate within a mix of formal regulations and individual discretion.

### *5.3. Practical Implications*

The findings offer several practical implications for healthcare administrators and policymakers. First, hospitals seeking to enhance environmental sustainability should invest in developing green transformational leadership capabilities among nurse managers. Training programs could focus on helping managers articulate compelling environmental visions, model sustainable behaviors consistently, and stimulate creative thinking about green solutions in clinical practice. Given the large effect size observed for green transformational leadership, such investments are likely to yield meaningful returns in terms of improved sustainability outcomes.

Second, healthcare organizations should attend to the ethical climate of the workplace as a foundation for sustainability initiatives. Climates that emphasize caring for all stakeholders, adherence to professional standards, and consistent application of institu-tional rules appear to create fertile ground for environmental responsibility. Hospital administrators can reinforce ethical climates through explicit communication of values, recognition of ethical conduct, and ensuring that policies and practices align with stated principles.

Third, the significant mediating role of green psychological climate suggests that hospitals should work to embed environmental values throughout the organizational culture. This can be achieved through visible sustainability policies, green procurement practices, environmental training programs, and recognition systems that reward sustainable behaviors. When nurses perceive that both the organization and their colleagues genuinely value environmental responsibility, they are more likely to internalize these values and act accordingly.

Fourth, and perhaps most critically, the findings regarding perceived organizational hypocrisy underscore the importance of authenticity in sustainability communications and initiatives. Healthcare leaders must ensure that their actions align with their words regarding environmental responsibility. This means allocating adequate resources for sustainability initiatives, personally modeling green behaviors, and avoiding symbolic commitments that are not backed by genuine effort. The data suggest that inconsistency between rhetoric and action not only fails to promote sustainability but may actually undermine the positive effects of leadership and ethical climate. Hospitals should conduct regular assessments of employee perceptions regarding organizational authenticity and address any gaps between stated values and actual practices.

Fifth, for policymakers in developing countries like Jordan, these findings suggest that behavioral interventions targeting leadership and organizational climate may complement infrastructural investments in achieving healthcare sustainability goals. While technological solutions such as energy-efficient equipment and waste management systems are necessary, their effectiveness depends partly on the willingness of frontline healthcare workers to adopt sustainable practices. Creating supportive organizational environments through leadership development and ethical climate enhancement can maximize the return on infrastructural investments.

### *5.4. Limitations*

Despite its contributions, this study has several limitations that should be acknowledged. First, the sample was drawn exclusively from Jordanian hospitals, which may limit generalizability to other national and cultural contexts. While Jordan provides a valuable setting for studying sustainability under resource constraints, future research should replicate these findings in other Middle Eastern countries and in diverse healthcare systems globally to assess the cross-cultural validity of the proposed model.

Second, the study focused on nurses as the target population. While nurses represent the largest segment of the healthcare workforce and have direct responsibility for many sustainability-relevant practices, future research should examine whether similar patterns hold for other healthcare professionals, including physicians, administrative staff, and support personnel.

Third, although the operationalization of perceived organizational hypocrisy at the supervisor level was theoretically grounded in social information processing theory, this measurement choice has implications for interpretation. The four items exclusively capture perceptions of the nurse manager's behavioral consistency regarding sustainability and do not reflect hypocrisy perceptions arising from other sources, such as discrepancies between hospital-wide sustainability communications and actual resource allocation or between senior leadership's stated commitments and institutional priorities. The moderating effects reported in this study should therefore be interpreted as reflecting supervi-sor-level hypocrisy specifically and may underestimate the full influence of perceived organizational inconsistency on sustainable behaviors. Future research should develop multi-level measures that distinguish hypocrisy perceptions arising from supervisors, units, and hospital administration.

Finally, while this study examined green psychological climate as a mediator and perceived organizational hypocrisy as a moderator, other mechanisms and boundary conditions warrant investigation. For instance, individual differences such as personal environmental values, green self-efficacy, or professional identity strength may moderate responses to organizational sustainability initiatives. Similarly, team-level factors such as workgroup norms or peer influence processes may play important roles in shaping individual sustainable behaviors.

## 6. Conclusions

This study investigated the organizational determinants of sustainable clinical behaviors among registered nurses in Jordanian hospitals. Drawing on social information processing theory and social learning theory, the research examined how green transformational leadership and ethical climate influence nurses' engagement in environmentally sustainable practices, the mediating role of green psychological climate, and the moderating role of perceived organizational hypocrisy.

The findings confirmed all six hypotheses, revealing that both green transformational leadership and ethical climate positively predict sustainable clinical behaviors among nurses. Green psychological climate emerged as a significant mediating mechanism, indicating that organizational signals about sustainability are translated into individual behaviors through employees' perceptions of the workplace environmental orientation. Importantly, perceived organizational hypocrisy was found to weaken the positive effects of both leadership and ethical climate on sustainable behaviors, highlighting that authenticity and consistency between words and actions are essential for effective sustainability initiatives.

These findings carry important implications for healthcare organizations seeking to reduce their environmental footprint. Hospital administrators should invest in developing green transformational leadership capabilities among nurse managers, cultivate ethical climates that emphasize stakeholder welfare and professional standards, and work to embed environmental values throughout the organizational culture. Critically, leaders must ensure that their sustainability communications are backed by genuine commitment and consistent action, as perceived hypocrisy can undermine otherwise effective organizational efforts.

The study contributes to the literature by extending green transformational leadership theory to the healthcare sector, integrating ethical climate research with environmental sustainability, introducing perceived organizational hypocrisy as a boundary condition in healthcare sustainability research, and providing evidence from a developing country context where resource constraints create both economic and ecological imperatives for conservation. By demonstrating that nurses' sustainable behaviors are shaped by leadership, ethical climate, and perceptions of organizational authenticity, this research offers a foundation for evidence-based interventions aimed at greening healthcare delivery.

As healthcare systems worldwide grapple with the paradox of promoting health while contributing to environmental degradation, understanding the organizational factors that enable frontline workers to adopt sustainable practices becomes increasingly critical. This study suggests that fostering sustainable clinical behaviors requires not only supportive leadership and ethical organizational environments but also genuine commitment that nurses can observe and trust. Future research should continue to explore the interplay between positive organizational antecedents and potential undermining factors across diverse healthcare contexts to advance both theory and practice in healthcare sustainability.

**Author Contributions:** Conceptualization: S.N. and T.A., Methodology: S.N., Data Collection: T.A., Data Curation: T.A., Formal Data Analysis: S.N. and T.A., Writing – original draft: S.N. and T.A., Writing – review and editing: S.N..

**Funding:** This work was supported by the Deanship of Scientific Research, Vice Presidency for Graduate Studies and Scientific Research, King Faisal University, Saudi Arabia KFU260785.

**Institutional Review Board Statement:** This study was approved by the Research Ethics Committee at King Faisal University (Approval No. KFU-2026-ETHICS3976).

**Data Availability Statement:** The data that support the findings of this study are available from the corresponding author upon reasonable request.

**Conflicts of Interest:** The authors declare no conflicts of interest.

## Appendix A

Table A1. Measurement Items for Study Variables.

| Variable | Items | Source |
| --- | --- | --- |
| Green Transformational Leadership | (1) My nurse manager inspires staff with environmental plans for the unit.<br>(2) My nurse manager provides a clear environmental vision for staff to follow.<br>(3) My nurse manager gets staff to work together for the same environmental goals.<br>(4) My nurse manager encourages staff to achieve environmental goals.<br>(5) My nurse manager acts with consideration of the environmental beliefs of staff.<br>(6) My nurse manager stimulates staff to think about green ideas. | [18] |
| Ethical Climate (Caring) | (1) What is best for everyone in the hospital is the major consideration here.<br>(2) The most important concern is the good of all the people in the hospital as a whole.<br>(3) Our major concern is always what is best for the patient.<br>(4) In this hospital, people look out for each other's good.<br>(5) It is expected that you will always do what is right for the patients and public. | [16] |
| Ethical Climate (Law and Code) | (1) People are expected to comply with the law and professional nursing standards over and above other considerations.<br>(2) In this hospital, the law or ethical code of the nursing profession is the major consideration.<br>(3) In this hospital, people are expected to strictly follow legal or professional standards.<br>(4) In this hospital, the first consideration is whether a decision violates any law or professional code. | [16] |
| Ethical Climate (Rules) | (1) It is very important to follow the hospital's rules and procedures here.<br>(2) Everyone is expected to stick by hospital rules and procedures.<br>(3) People in this hospital strictly obey the hospital policies. | [16] |
| Green Psychological Climate (Organization) | (1) Our hospital is worried about its environmental impact. (2) Our hospital is interested in supporting environmental causes. (3) Our hospital believes it is important to protect the environment. (4) Our hospital is concerned with becoming more environmentally friendly. | [17] |
| Green Psychological Climate (Co-workers) | (1) In our hospital, employees pay attention to environmental issues.(2) In our hospital, employees are concerned about acting in environmentally friendly ways.(3) In our hospital, employees try to minimize harm to the environment.(4) In our hospital, employees care about the environment. | [17] |
| Perceived Organizational Hypocrisy | (1) My nurse manager tells us to follow environmental rules but does not follow them him/herself.(2) I wish my nurse manager would practice what he/she preaches about sustainability more often.(3) My nurse manager says one thing about environmental practices but does another.(4) There is a gap between what my nurse manager says and does regarding sustainability. | Adapted from [37] |
| Sustainable Clinical Behaviors (Conservation) | (1) I turn off lights in patient rooms and work areas when not in use.(2) I switch off medical equipment and computers when not needed.(3) I avoid unnecessary water usage during clinical procedures.(4) I print documents double-sided whenever possible.(5) I use digital documentation instead of paper when feasible. | Developed for this study based on [34] |
| Sustainable Clinical Behaviors (Waste Management) | (1) I properly segregate medical waste from general waste.(2) I dispose of recyclable materials in designated bins.(3) I avoid opening unnecessary supplies that may go unused.(4) I return unused or unopened supplies to storage rather than discarding them. | Developed for this study based on [34] |
| Sustainable Clinical Behaviors (Proactive) | (1) I suggest environmentally friendly practices to my supervisor or hospital committees.(2) I encourage colleagues to engage in sustainable practices at work.(3) I participate in hospital sustainability programs or initiatives when available.(4) I bring reusable items to work (e.g., water bottle, food containers). | Developed for this study based on [34] |